\documentclass{sigchi}

\toappear{\scriptsize Permission to make digital or hard copies of all or part of this work for personal or classroom use is granted without fee provided that copies are not made or distributed for profit or commercial advantage and that copies bear this notice and the full citation on the first page. Copyrights for components of this work owned by others than the author(s) must be honored. Abstracting with credit is permitted. To copy otherwise, or republish, to post on servers or to redistribute to lists, requires prior specific permission and/or a fee. Request permissions from permissions@acm.org. \\
{\emph{DIS '20, July 6--10, 2020, Eindhoven, Netherlands.} } \\
\copyright~2020 Copyright is held by the owner/author(s). Publication rights licensed to ACM. \\
ACM ISBN 978-1-4503-6974-9/20/07\ ...\$15.00.\\
https://doi.org/10.1145/3357236.3395528
}

\usepackage{color-edits}
\addauthor{sw}{blue}
\addauthor{haiyi}{red}
\addauthor{loren}{darkpastelgreen}
\addauthor{bowen}{orange}
\addauthor{irene}{yellow}
\addauthor{jodi}{purple}

\usepackage{multirow}
\usepackage{balance}       
\usepackage{graphics}      
\usepackage[T1]{fontenc}   
\usepackage{txfonts}
\usepackage{mathptmx}
\usepackage[pdflang={en-US},pdftex]{hyperref}
\usepackage{color}
\usepackage{booktabs}
\usepackage{textcomp}
\usepackage{amsmath}
\usepackage[toc,page]{appendix}

\usepackage{microtype}        
\usepackage{ccicons}          

\usepackage{todonotes}

\def\plaintitle{Keeping Designers in the Loop: Communicating Inherent Algorithmic Trade-offs Across Multiple Objectives }

\def\emptyauthor{}
\def\plainkeywords{Interactive visualization; algorithmic fairness; algorithmic trade-offs; criminal prediction; case study; experimental design; interview study.}

\makeatletter
\def\url@leostyle{%
  \@ifundefined{selectfont}{
    \def\UrlFont{\sf}
  }{
    \def\UrlFont{\small\bf\ttfamily}
  }}
\makeatother
\def\pprw{8.5in}
\def\pprh{11in}

\definecolor{darkpastelgreen}{rgb}{0.01, 0.75, 0.24}
\definecolor{britishracinggreen}{rgb}{0.0, 0.26, 0.15}
\definecolor{darkgreen}{rgb}{0.0, 0.2, 0.13}
\definecolor{linkColor}{RGB}{6,125,233}
\hypersetup{%
  pdftitle={\plaintitle},
  pdfauthor={\emptyauthor},
  pdfkeywords={\plainkeywords},
  pdfdisplaydoctitle=true, 
  bookmarksnumbered,
  pdfstartview={FitH},
  colorlinks,
  citecolor=black,
  filecolor=black,
  linkcolor=black,
  urlcolor=linkColor,
  breaklinks=true,
  hypertexnames=false
}

\begin{document}

\title{\plaintitle}

 \numberofauthors{1}
 \author{%
   \alignauthor{Bowen Yu\textsuperscript{1}, Ye Yuan\textsuperscript{1}, Loren Terveen\textsuperscript{1}, Zhiwei Steven Wu\textsuperscript{1}, Jodi Forlizzi\textsuperscript{2},   Haiyi Zhu\textsuperscript{2}\\
   \affaddr{\textsuperscript{1}University of Minnesota - Twin Cities, 
   \textsuperscript{2}Carnegie Mellon University} \\
   \email{\textsuperscript{1}\{bowen-yu, yuan0191, terveen,
 zsw\}@umn.edu, \textsuperscript{2}\{forlizzi, haiyiz\}@cs.cmu.edu}}\\
 }




\maketitle

\begin{abstract}
Artificial intelligence algorithms have been used to enhance a wide variety of products and services, including assisting human decision making in high-stake contexts. However, these algorithms are complex and have trade-offs, notably between prediction accuracy and fairness to population subgroups. This makes it hard for designers to understand algorithms and design products or services in a way that respects users' goals, values, and needs. We proposed a method to help designers and users explore algorithms, visualize their trade-offs, and select algorithms with trade-offs consistent with their goals and needs. We evaluated our method on the problem of predicting criminal defendants' likelihood to re-offend  through (i) a large-scale Amazon Mechanical Turk experiment, and (ii) in-depth interviews with domain experts. Our evaluations show that our method can help designers and users of these systems better understand and navigate algorithmic trade-offs. This paper contributes a new way of providing designers the ability to understand and control the outcomes of algorithmic systems they are creating.
\end{abstract}

\ccsdesc[300]{Human-centered computing~Interactive systems and tools}



\keywords{\plainkeywords}

\printccsdesc

\section{Introduction}

Artificial Intelligence algorithms are being used to support and enhance a wide variety of products and services, often with critical impacts on people's lives. Examples include (i) helping judges decide whether criminal defendants should be detained or released while awaiting trial \cite{dressel2018accuracy, corbett2017algorithmic}, (ii) assisting child protection agencies in screening referral calls \cite{chouldechova2018case}, (iii) helping employers filter job resumes \cite{weber2012your}, and (iv) facial recognition, which can be used for surveillance and crime prevention \cite{buolamwini2018gender}. On social media sites such as Facebook, algorithms are used to identify and censor trolls, fake news, terrorism, racist and sexist ads. 





One outstanding design challenge in integrating AI capabilities into real-world applications is the communication gap between designers, users, and algorithm developers. 
Designers (who design products or service that use AI algorithms) and users (who directly interact with or are affected by these products and services) know their goals and needs, but struggle to understand AI capabilities and envision how to design an algorithm to achieve their goals \cite{yang2020}. On the other hand, algorithm developers know a lot about how to create algorithms and tune the algorithms to optimize for certain system criteria, but know little about how these algorithmic choices can influence the user experience. There is a need for tools to facilitate the communication of usability considerations in algorithm implementation between designers, users, and algorithm developers.

In practice, some experienced designers have started to work closely with algorithm developers to identify design goals that are both technically viable and improve the user’s experience\cite{yang2018investigating}.
However, merely identifying design goals and mapping them to algorithmic criteria is not sufficient. There are often \textit{\textbf{inherent trade-offs in implementing multiple design goals in the algorithm}}. Optimizing for multiple criteria is challenging: optimizing one criterion often leads to poor performance on others. For example, when developing a risk assessment tool in order to aid judges' decisions on detaining or releasing defendants while awaiting trial, designers and users (judges) may have dual goals: (1) not detaining someone who will not re-offend, and (2) not releasing someone who will re-offend. The two goals correspond to two system criteria: reducing \textit{false positives} 
and reducing \textit{false negatives} 
in the predictive model. However, there is a well-documented trade-off between false positive and false negative: reducing false positives can increase false negatives and vice versa \cite{klinkman1998false, dove2017ux}. Furthermore, judges may want the tools to make predictions that are both \textit{accurate} in general and \textit{fair} to defendants across different demographic groups. However, machine learning research has shown a trade-off between prediction accuracy and fairness \cite{kleinberg2018algorithmic, agarwal2018reductions, menon2018cost, kearns2019empirical, dwork2018decoupled}. Specifically, improving fairness -- such as minimizing differences in false-positive rates between different racial or gender groups -- can lead to a decrease in overall prediction accuracy.

Algorithmic trade-offs are critical: they impact the intended user experience, and sometimes even raise serious ethical concerns or result in societal-level consequences.
The HCI community has recognized that designers and users struggle to work with AI, and has proposed design processes and methods to help address the challenges (e.g., \cite{yang2020, amershi2019guidelines, van2018prototyping,cheng2019explaining}). 
However, to the best of our knowledge, few studies have investigated techniques for explicitly communicating and explaining the \textit{trade-offs} in implementing multiple design objectives in the algorithm. Our research takes on this challenge.

We propose the following method to communicate the trade-offs between multiple design objectives in AI prediction algorithms. First, given a set of design objectives (and corresponding system criteria), generate a family of prediction models with a wide spectrum of trade-offs. Second, create interactive interfaces to visualize the trade-offs. The interfaces should allow designers and users to explore the trade-offs between the family of models. The goal is to help them select specific models that are consistent with their needs and values.

We conducted a case study in the context of \textit{recidivism} prediction (predicting whether or not a defendant will re-offend) to illustrate the method. We chose this context  because: (i) as noted above, this is a high-stakes decision, and (ii) the machine learning community has intensely studied trade-offs between different accuracy and fairness notions in this context \cite{andrews2006recent, singh2014international, skeem2016risk}.  
To evaluate the effectiveness in communicating the algorithmic trade-offs, 
we conducted two studies: (i) a large-scale Amazon Mechanical Turk experiment, and (ii) in-depth interview sessions with domain experts.
We found that (i) we effectively communicated algorithm trade-offs and significantly improve non-algorithm-expert participants' understanding of algorithmic trade-offs; (ii)  participants were able to navigate between a wide range of machine learning models and select a model with their most acceptable trade-offs. 

Our case study also suggested some unintended consequences of making algorithmic trade-off transparent. First, we observed great diversity in model selection among our participants, which suggests future research opportunities for creating mechanisms to enable trade-off discussion and negotiation. Second, we found that communicating algorithmic trade-offs also affected participants' trust in AI supported decision making in general. Almost 50\% of participants changed self-report trust in prediction algorithms after using our interfaces, with some increasing their trust and others decreasing. 

Our findings show that our method can help designers and users of these systems better understand algorithmic trade-offs. They can explore different possibilities in the vast design space of all algorithmic possibilities. As such, this paper contributes a new way for communicating how algorithms work, and for giving designers and users ability to understand and control the outcomes of algorithmic systems they are creating.

\section{Related Work}

\subsection{Challenges of Working with AI}

The HCI community has recognized that designers and developers struggle to innovate with AI and Machine Learning techniques. Dove et al. \cite{dove2017ux} conducted surveys with UX design professionals about the challenges they face when working with AI. Their findings showed that many UX designers struggle to understand the capabilities and limitations of AI, and they typically joined projects towards the end, after the functional and algorithmic decisions had been made \cite{dove2017ux}. Yang et al. \cite{yang2020} further synthesized prior work and their own experience, and identified two unique challenges of envisioning and prototyping with AI: (1) the uncertainty surrounding
AI’s capabilities and (2) AI’s output complexity.

Several different approaches to helping designers and developers innovate with AI have emerged. We next review research aimed at improving the general "explainability" of AI, and then discuss work that develops tools, techniques, and features to help designers work with AI. 





\subsection{Explaining AI and Machine Learning Algorithms}
Several communities are paying attention to the issue of explaining how AI (and particularly machine learning) works.

The explainable artificial intelligence (XAI) community (see \cite{biran2017explanation} for a review) aims to provide users with explanations of algorithms' decisions in some level of detail to ensure that the algorithms perform as expected \cite{gilpin2018explaining}. Researchers have made progress on transforming complex models, such as neural networks into simple ones (such as linear models or decision trees), through approximation of the entire model \cite{craven1999rule}  
or local approximation \cite{ribeiro2016model}. 
Visualization techniques have been developed to explain different types of machine learning models. Examples include traditional machine learning models such as linear models \cite{ribeiro2016should}, decision trees \cite{lakkaraju2016interpretable}, and ensemble classifiers \cite{wyner2017explaining}, and deep neural networks \cite{gilpin2018explaining}.


The HCI community aims to improve the usability of explanation interfaces through user centered design and evaluation approaches. For example, Carter and Nielsen\cite{carter2017using} created user interfaces which explains the representations inside machine learning models, and give people new tools for reasoning. Krause et al\cite{krause2016interacting} developed \textit{Prospector} to provide interactive partial dependence diagnostics, which can help people understand how features affect the prediction overall.  Cheng et al. further conducted human-centered design and empirical evaluation of parallel interface prototypes to explore the effectiveness of different strategies (e.g., ``black-box'' versus ``white-box'', and ``interactive'' versus ``static'') to help non-expert stakeholders understand algorithmic decision making \cite{cheng2019explaining}.
The survey paper\cite{abdul2018trends} summarizes the field and outlines a blueprint for CHI research on explainable AI. 




Most of this work focuses on improving the explainability of AI systems' individual decisions. However, a recent interview study with pathologists about a diagnostic AI assistant found that users also wanted to know the design objectives of the AI systems and the "inherent trade-offs that the designers of the intelligent systems must navigate in implementing the system"  (\cite{cai2019hello}), which opens new challenges and motivates our work. 

\subsection{Tools, Techniques and Features to Help Work with AI}

Researchers have developed tools, methods, abstractions, exemplars, and guidelines to support designers in working with algorithm developers during all phases of the design process. 

Researchers have developed tools for designers to prototype AI and develop one’s own classifiers. For examples, van Allen\cite{van2018prototyping} developed the Delft AI Toolkit, with the goal of allowing designers to experiment and gain a stronger understanding of AI as a material of design. Fiebrink and Cook \cite{fiebrink2010wekinator} developed Wekinator, a system for using machine learning to build real-time interactive systems in music. 


Sets of design guidelines have been published to provide knowledge about how to display the outputs of AI systems to users \cite{appleAIguideline,googleAIguideline,amershi2019guidelines}. For example, Amershi et al \cite{amershi2019guidelines} proposed 18 generally applicable design guidelines for human-AI interaction. The guidelines provide useful resources for designers and design teams working with AI. 



Design features (e.g., abstractions and examplars, and design patterns) have been used to help designers innovating with AI. 
For example, Yang et al \cite{yang2018investigating} found that UX designers comprehend ML largely through abstractions and exemplars:"the abstractions served as a general insight about an ML capability and provided an understanding of how it worked. The design exemplars provided specific interaction possibilities and a glimpse of a possible felt experience". However, they also found that designers were confused by the published exemplars, and wanted to directly create their own exemplars instead of using pre-existing ones. 
Design patterns have been shown to be useful to help practicing designers understand how to work effectively in a new domain of design. For example, designers found design patterns useful to help understand the design space of privacy and security features in software\cite{chung2004development}. Researchers \cite{yang2016planning} also provided design patterns for designers to sketch through a mobile app and create predictive interfaces. They are useful for understanding how algorithms work in analogous design examples, such as recommendations. 

\subsection{Research Gap} 


In sum, prior research has been conducted to help designers understand how AI works, provide insights about AI capacities and possibilities, and suggest best practices and guidelines. However, little work has been conducted to help designers engage in the algorithm implementation and participate in the decision-making of algorithmic choices. In practice, once design goals are set, even experienced designers often do not participate in the algorithm implementation process, but only focus on crafting its interaction design \cite{yang2018investigating}. 

In this paper, we take on the challenge to facilitate communication of inherent algorithmic trade-offs. The goal is to involve designers and users in the algorithmic trade-off decision-making, which often have critical impacts on the intended user experience, and even societal-level consequences.

\section{Approach Overview: Communicate Inherent
Algorithmic Trade-offs}


In this paper, we propose a novel method to communicate the trade-offs in implementing multiple design objectives in the algorithm. The method contains two steps: generating a family of models on a wide spectrum of trade-offs, and visualizing the trade-offs between the models.  

Note that we assume that at this stage, designers and machine learning developers have already identified a set of design goals and have mapped them into system criteria. For example, major social networking sites increasingly rely on algorithmic tools to automatically identify and censor hate speech, adult content, misinformation, racist and sexist ads, and so forth. Designers who work with users and algorithm developers might identify the following design goals for the content moderation tools: (1) catching all the undesirable content (minimizing false negatives), (2) not falsely accusing any well-intentioned users (minimizing false positives), and (3) being fair to users from different demographic groups (equalize false positives/negatives). With the set of design objectives (and associated system criteria), we step through the following process to communicate the algorithmic trade-offs. 


\textbf{Step 1: Generate a family of models to capture the trade-offs}. Given a collection of system criteria that correspond to design goals, the first step is to generate a family of predictive models that exhibit a wide range of trade-offs between the different system criteria. One technique we can use to capture the trade-offs is to identify \emph{Pareto-optimal} models (e.g., \cite{gerry, empirical_gerry} ). More formally, given a set of system criteria, we say that a model is \emph{Pareto-optimal} if there is no alternative model that is strictly better than the given model on all the system criteria. 

\textbf{Step 2: Make the trade-offs between models interpretable.} Given a family of models, the next step is to develop methods to communicate models’ trade-offs. For example, interactive visualizations can be used to enable users and designers to "play" with the models, understand and explore the trade-offs, and select a suitable model.

Below we describe a “proof-of-concept” case study in the context of recidivism prediction to illustrate our process in detail and to show its value and promise.








\section{Case Study: Recidivism Prediction}

We used recidivism prediction (predicting whether a defendant will or will not re-offend) as the context for exploring the general research problem of helping people understand intelligent algorithms and communicate their trade-offs.

\subsection{Context}
In 2016, ProPublica, an independent, non-profit newsroom that produces investigative journalism in the public interest, published reports about COMPAS - a case management and decision support tool used by U.S. courts to assess the likelihood of a defendant re-offending\cite{HowWeAnalyzedCOMPAS}. The analysis shows that African American defendants were much more likely to be misclassified as high risk to re-offend compared to their white counterparts \cite{MachineBias, HowWeAnalyzedCOMPAS}.  

Their findings led to a growing body of research aiming at integrating fairness into machine learning, often referred to as fairness-aware machine learning.  A lot of the work aims to formulate fairness notions as algorithmic constraints and build predictive models that satisfy fairness notions, including statistical parity \cite{dwork2018decoupled}, equalized opportunity \cite{hardt2016equality}, and calibration \cite{pleiss2017fairness}. However, for many of these fairness measures, prior research has identified a range of trade-offs between fairness and accuracy \cite{agarwal2018reductions, kearns2019empirical, menon2018cost, dwork2018decoupled}. Recent studies indicate that different desirable notions of fairness are not only incompatible with each other \cite{pleiss2017fairness}, but  often mutually exclusive \cite{KMR16, Chouldechova17}. 

Despite the mathematical rigor of these approaches, a recent interview study with 27 public sector ML practitioners suggested a disconnect between the current fairness-aware machine learning research and users' and stakeholders’ realities, context, and constraints; this disconnect is likely to undermine practical initiatives \cite{veale2018fairness}. To address the challenge, prior work suggests the importance of engaging users and stakeholders throughout the algorithm design process \cite{lee2019procedural, zhu2018value}.

\subsection{Data}
We recreated a recidivism prediction tool using a data set provided by ProPublica\cite{HowWeAnalyzedCOMPAS}. The dataset originally contains information of 11,757 defendants including their prior criminal history, jail and prison time, and demographics (such as race, gender, and age) \cite{dieterich2016compas}. 
We followed the literature to formulate the problem as binary classification, and the labels are whether a defendant commits ``a new misdemeanor or felony offense within two years of the COMPAS administration date'' \cite{HowWeAnalyzedCOMPAS}. 
We removed defendants whose records were not complete and who were just charged for traffic offenses and municipal ordinance violations. 
For the purpose of the case study, we only focused on two protected attributes, race and gender, and constrained them to be binary, e.g., African American and White , female and male.
To better illustrate trade-offs, we created two balanced data sets for race and gender separately. This resulted a data set of 3,000 defendants (1,500 White defendants and 1,500 African American defendants) and a data set of 1,600 defendants (800 male defendants and 800 female defendants). We ran logistic regression on the two data sets. The prediction accuracy for the two data sets is 0.715 and 0.721, respectively, with a random 70\%-30\% split on train and test data, which is consistent with results of previous studies.

\subsection{Design Objectives and System Criteria}

Based on the prior literature, we defined the following major design goals for the recidivism prediction tool: 

\begin{itemize} 
    \item\underline{\textit{Not falsely detain defendants who will not re-offend.}} This is unfair to the defendants and costly to society. This goal corresponds to reducing "false positives".
    \item\underline{\textit{Not release defendants who will re-offend.}} This is dangerous for society. This goal corresponds to reducing "false negatives". 
    \item \underline{\textit{Be fair to defendants across different demographic groups.}} We define (un)fairness as disparity in false positive and false negative rates between different groups. However, prior research shows that equalizing false positives and false negatives between different groups might increase the overall error rate, which is undesirable~\cite{agarwal2018reductions,kearns2019empirical}.
\end{itemize}

\begin{figure}[!ht]
\centering
\includegraphics[width=1.0\columnwidth]{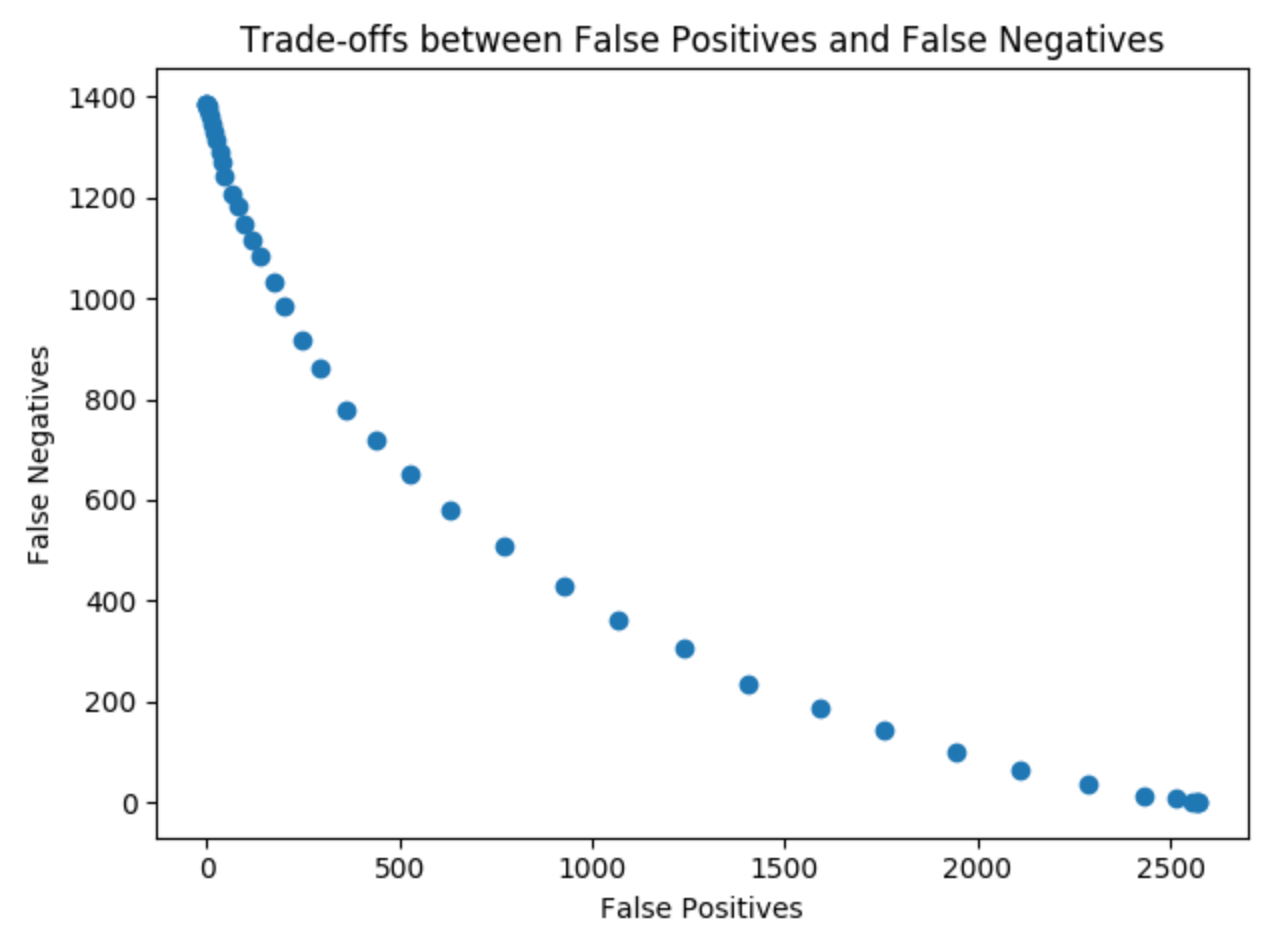}
  \caption{Relationship between false positives and false negatives.}
  \label{fig:tradeoffs_fn_fp}
\end{figure}
\begin{figure}[!ht]
\centering
\includegraphics[width=1.0\columnwidth]{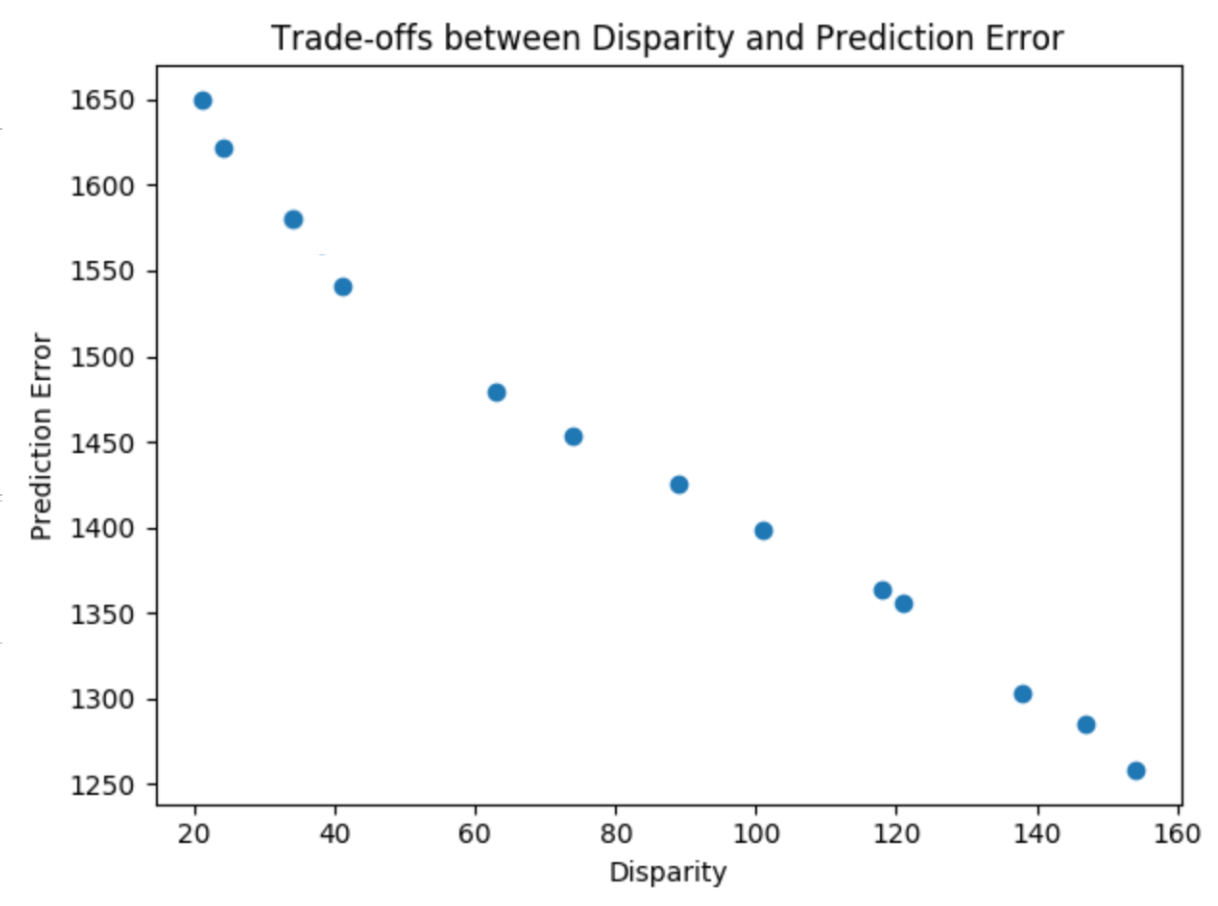}
  \caption{Relationship between overall prediction errors and disparity between different racial groups, with threshold 0.45.}
  \label{fig:tradeoffs_disp_error}
\end{figure}

\subsection{Step 1: Generate a Family of Models}

We now discuss how we generated a set of models with different trade-offs across a variety of system criteria.

\subsubsection{Trade-Offs between False Positives and False Negatives}

To capture the trade-off between false positives and false negatives, we varied a classification threshold. We first had to define this threshold, which required mapping a probability to a binary category, where a value above the threshold indicates ``re-offend'', and a value below indicates ``not re-offend''. 
Figure \ref{fig:tradeoffs_fn_fp} shows the relationship between false positives and false negatives when we varied the  threshold in our data. 




\subsubsection{Trade-Offs between Overall Errors and Disparity (Unfairness).}

We followed the prevalent statistical fairness approach in the machine learning fairness literature. We selected a small number of groups specified by sensitive attributes, and then sought approximate equality of these groups on certain statistics of the predictor, such as false-positive rates and false-negative rates. In our study, we considered two groups, denoted by  $a_0$ and $a_1$, and specified by either race or gender. We then formulated our (un)fairness measure as the \emph{disparity} between the number of false positives $FP$ and false negatives $FN$ between the two groups:
\begin{equation}\label{formal}
    \max(|FP(a_1) - FP(a_0)|, |FN(a_1) - FN(a_0)|).
\end{equation}

Note that we chose to use counts instead of ratios, since counts are easier to explain to non-expert stakeholders.
To capture trade-offs between  overall accuracy and disparity, we adapted the algorithms from~\cite{agarwal2018reductions, gerry} to generate the set of \emph{Pareto-optimal} predictive models, for which it is impossible to improve either criterion without worsening the other. Figure \ref{fig:tradeoffs_disp_error} shows a Pareto curve of prediction errors and disparity on African American and White defendants. Models on the right side of Figure \ref{fig:tradeoffs_disp_error} prioritize ``minimizing overall errors'', while models on the left side prioritize ``minimizing disparity''. We can observe that by reducing the disparity between the two groups from 158 to 21, overall prediction errors increase from 1253 to 1651. The technical details on how we generated the Pareto curves are included in the supplementary materials.



\subsection{Step 2: Make the Trade-Offs Interpretable}





We want to develop interactive interfaces to let users explore and compare a set of prediction models with a spectrum of trade-offs between false negatives, false positives, overall prediction errors and fairness measures. Since our tool is designed for designers and users without technical background, our design must be able to effectively communicate relevant technical features of the model to a non-technical audience.

In this case study, we experimented with two interface strategies to visualize and explore the models: \textit{confusion matrix} and \textit{text}. A confusion matrix is a common approach to visualize model performance in machine learning  \cite{talbot2009ensemblematrix}; it displays four quadrants representing four types of prediction outcomes (false positives, false negatives, true positives, and true negatives).  We also can communicate model performance simply through textual explanations.

\subsection{Interface Designs}

We followed a human-centered design process \cite{holtzblatt2004rapid} involving a number of iterations. We started with a brainstorming session to ideate different design directions and features based on our design requirements. Next we synthesized and clustered the ideas. We then incorporated the ideas into the creation of low-fidelity prototypes, and conducted informal qualitative analysis and pilot studies to evaluate and improve the prototypes. Each step in this process provided rich insights from users' perceptions and helped to shape our final design and implementation.

Our final interface design consists of a two-part layout: (i) a control panel that lets users select models, and (ii) a result panel that shows relevant results of the selected model. The results panel is based on one of our two model visualization strategies, the confusion matrix view (Figure \ref{fig:overal_data}) or the text view (Figure \ref{fig:overal_scenario}). 

We now describe our interface in our detail.

\subsubsection{Control Panel}
Since we had two types of trade-offs to communicate to users, we designed two separate controls. 

For  trade-offs between false positives and false negatives (See the upper section of the control panel in Figure \ref{fig:overal_data}), we designed a control bar that lets users adjust the threshold. 

For trade-offs between errors and disparity (see the lower section of the control panel in Figure \ref{fig:overal_data}), once a protected attribute (gender or race) is selected, we present the Pareto curve with the given threshold. We let users select any particular model shown on the Pareto curves.

\subsubsection{Result Panel}
Prior work has compared the use of visualizations and texts for communicating various statistical aspects of algorithms \cite{gleicher2011visual}. Research shows that visualizations are more effective at grabbing user attention, but studies also reported that some users may prefer text over visual content. Therefore, we decided to design two different views in the result panel.

\begin{figure*}[!ht]
  \centering
  \includegraphics[width=2.0\columnwidth]{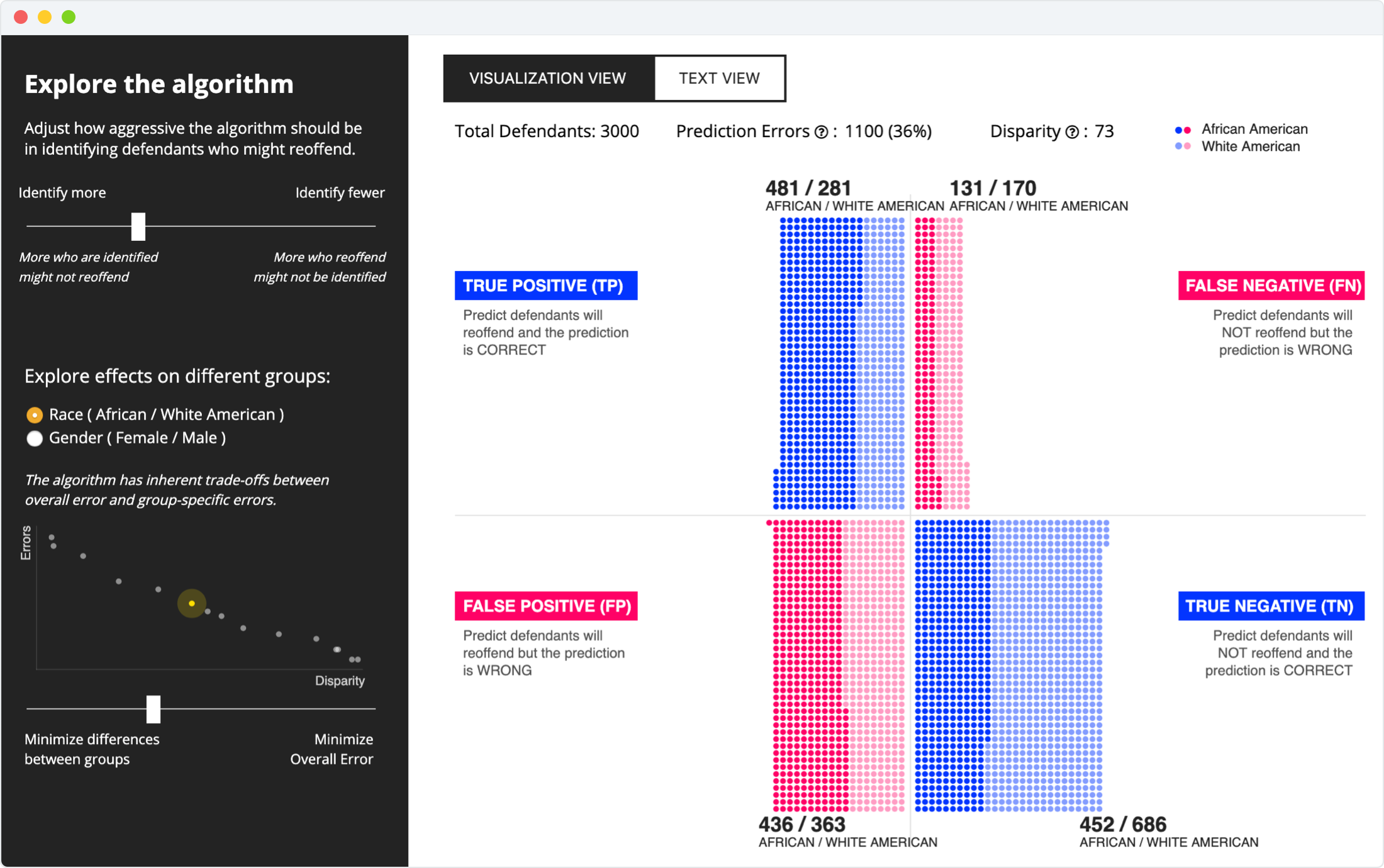}
  \caption{
  The confusion matrix view of the interface. The left-hand side is the control panel that provides control options for users to adjust trade-offs between the two types of errors and trade-offs between prediction errors and disparity. The right-hand side is the result panel that presents the overall prediction errors and disparity, and the four types of prediction outcomes in four quadrants with a short text explanation of the terminology.
  }
  \label{fig:overal_data}


\includegraphics[width=2.0\columnwidth]{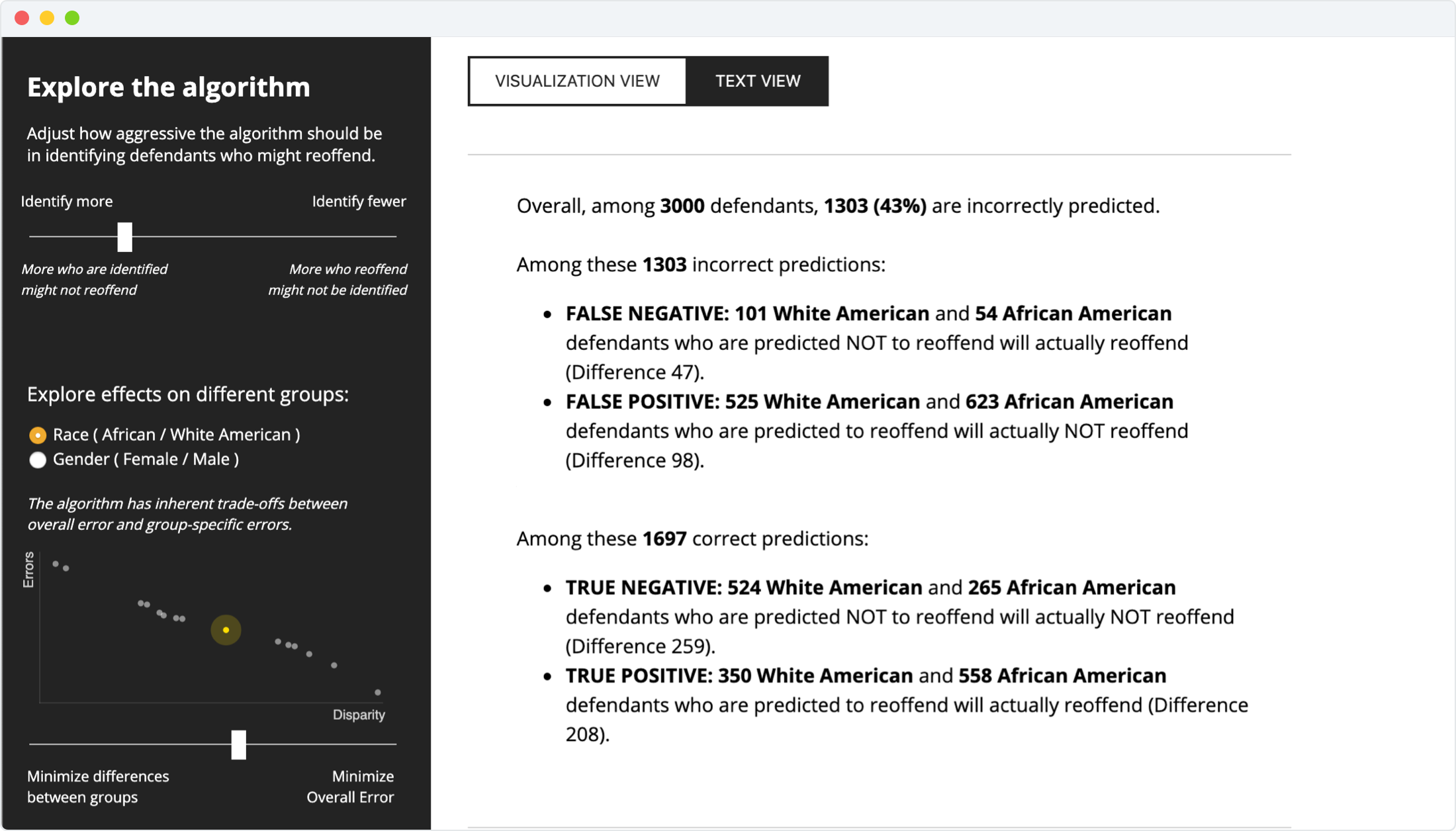}
  \caption{
  The text view of the interface. The left-hand side is the same control panel as the confusion matrix view that provides control options for two types of algorithmic trade-offs. The right-hand side is the result panel that presents the overall prediction errors and disparity and the four types of prediction outcomes in text description, separated by correct and incorrect predictions.
  }
  \label{fig:overal_scenario}
\end{figure*}

\textbf{\textit{Confusion Matrix View.}} 
 We created four separated quadrants with each dot representing one classified defendant in one of the four prediction categories (true positive, false positive, false negative, and true negative) (see the result panel in Figure \ref{fig:overal_data}). We also displayed the total number of defendants in each category. To distinguish correct and incorrect predictions, we applied two colors, blue and red, to highlight the difference. When a protected attribute (race or gender) is selected, the dots in each quadrant will be split into two different colors representing two groups under the selected protected attribute, such as African American and White defendants. As users move the control bars, the interface will display the changes in prediction outcomes accordingly. 

In addition, the interface provides the summary of the key metrics (e.g.,  prediction errors and disparity) on the top of the result panel. Explanations about the metrics will show up when users hover over the question marks next to the metrics.

\textbf{\textit{Text View.}} The text view (Figure \ref{fig:overal_scenario}) displays the same set of information as the confusion matrix view, but in plain text.  It describes the four prediction categories and states the number of defendants in each category. We followed a natural logic by grouping the prediction outcomes by incorrect or correct predictions (see the result panel in Figure \ref{fig:overal_scenario}). When a protected attribute (race or gender) is selected, we show the number for each group. Information about prediction errors and disparity is also described in text.

\section{Evaluation Overview}


We evaluated whether our method helps people comprehend, navigate, and manage trade-offs through (i) a large-scale Amazon Mechanical Turk experiment, and (ii) six in-depth interviews with domain experts.

Note that in this proof-of-concept study, we did not evaluate with UX designers, which we will discuss later in the paper as a limitation and opportunity for future work. However, UX designers focus on users. UX Designers conduct user research and evaluation to incorporate users' needs and preferences in the design of the products and services. 
The goal of our evaluation is to show our approach can help both novice users (AMT participants) and domain experts to understand and navigate algorithmic trade-offs, and express their preferences regarding the algorithmic trade-offs, which allows UX designers to design AI-based applications for the users. 

The goal of the Amazon Mechanical Turk experiment is to answer the following questions:

\begin{itemize}
  \item Q1: Can we help non-expert participants comprehend trade-offs in recidivism prediction?
  \item Q2: Are there differences between the confusion matrix-based interface and the text-based interface?
  \item Q3: Can our interfaces help participants navigate trade-offs and select models?
  \item Q4: Are there unintended consequences of making trade-offs transparent?
\end{itemize}

The interviews aim to explore how real users of recidivism prediction tools (e.g., judges, lawyers and policymakers) would use our interfaces. 

\section{Evaluation 1: Amazon MTurk Experiment}



\subsection{Experimental Design}



We conducted a randomized between-subjects experiment with three conditions: (i) \textit{confusion matrix view condition}:  participants used our confusion matrix view interface (Figure \ref{fig:overal_data}); (ii) \textit{text view}: participants used our text view interface (Figure \ref{fig:overal_scenario}); and (iii) \textit{baseline condition}: participants did not use \textit{any} interface, but instead proceeded directly to the questionnaire. We included the final condition to assess people's baseline understandings of trade-offs in machine learning. Participants were randomly assigned to one of the three conditions, and were allowed to spend as much time as needed to finish the evaluation questionnaires with or without the help of any interface.



\subsection{Participant Recruitment}
We recruited 301 participants from Amazon Mechanical Turk (AMT) in August 2019 for our study. To ensure the quality of survey responses, we recruited participants who had finished more than 100 tasks (HITs) with a task (HIT) approval rate of 95\% or above. We also ensured that participants were at least 18 years old and resided in the U.S., so that they had a higher chance of having contextual knowledge about the U.S. judicial system. 
The average time for completing the survey was 28.6 minutes. Each participant received a base payment of \$4 and an additional bonus (up to \$1.20) based on the number of correct answers they had in the objective comprehension questions (each correct answer would result in a bonus payment of \$0.20). To ensure participants would answer questions honestly without random guessing, we provide an \textit{``I don't know''} option for each question with a \$0.05 bonus payment. On average, each participant received a payment of \$8.70, which is higher than the minimum wage in the U.S. (\$7.25 per hour at the time of writing). 

301 participants finished the tasks on AMT, but 15 of them failed the attention check (not included in the analysis). Our analysis included 107 participants in the baseline condition, 93 participants in the confusion matrix view condition, and 86 participants in the text view condition. The demographic information including age, gender, race, education level, was comparable across the three conditions ($\chi^2$ not significant).

\subsection{Study Procedure}
Participants were informed that the purpose of the study was to help users understand intelligent algorithms that support people in making important decisions. We also designed quiz questions  to make sure participants understood the study context, such as the definition of \textit{recidivism} and \textit{prediction algorithms}. Participants had to answer the quiz questions correctly before they could proceed. We included the description of study context and quiz questions in the supplementary materials. 
Participants were given as much time as they wanted  to explore the interface and complete a set of questions (details will be described below in the ``Evaluation Metrics'' section). 
We also inserted an instructed-response question for an attention check, which directed respondents to choose a specific answer \cite{meade2012identifying, cheng2019explaining}.




\subsection{Evaluation Metrics}
We designed a set of questions to measure the following metrics.  
We include the questions in the supplementary materials.

\textit{\underline{Objective Comprehension.}}
Participants answered six multiple-choice questions with objectively correct answers to evaluate their understandings of the algorithms: four of them focused on understanding the basic concepts,  
while two of them assessed understanding of algorithmic trade-offs. 


\textit{\underline{Subjective Evaluation.} }
Participants self-reported how well they understood the algorithmic trade-offs on a Likert scale.


\textit{\underline{Model Selection.}} In the two interface conditions, participants were instructed to adjust the interface controls to select a model, and were asked questions about whether and why the selected model was most consistent with their values.

\textit{\underline{Trust.}}
We also measured participants' perceived trust of the algorithm's recidivism predictions. We adapted  questions from the prior literature that measured trust in human-machine systems on a 7-point Likert scale \cite{lee1992trust, corritore2003line, cheng2019explaining}. 
We asked the same set of questions about trust twice, both before and after participants explored our interfaces and/or answered the objective comprehension questions.





In addition, we also asked questions about participants' technical literacy and their demographic information (age, gender, and education levels). Technical literacy is measured using a 7-point Likert scale question to assess participants' familiarity with AI-powered systems \cite{wilkinson2009measurement, cheng2019explaining}.





\subsection{Results}
Our main findings include the following: 
\begin{itemize}
    \item Both the confusion matrix view and text view interfaces significantly improved participants' comprehension of algorithmic trade-offs compared to the baseline. There is no statistically significant difference between the confusion matrix view and the text view.  
    \item Our interfaces let participants select models that represented their values. We also observed great diversity in the selected models: some participants tended to balance trade-offs, while others concentrated on optimizing one criterion.
    \item Nearly half of the participants changed their trust in the algorithmic prediction after using our interfaces: \textit{22.3\%} trusted the algorithm \textit{more}, while \textit{25.1\%} trusted the algorithm \textit{less}.  
\end{itemize}

We next describe our analyses in more detail. 


\subsubsection{Improved Objective Comprehension of Trade-Offs}


We created a set of linear regression models (see Table \ref{tab:results}) with \textit{Objective Comprehension} and \textit{Subjective Evaluation} as dependent variables and experimental condition as the independent variable. Models 1, 2, and 3 in Table \ref{tab:results} show the differences between three experimental conditions (\textit{IsCMView} v.s. \textit{baseline}, and \textit{IsTextView} v.s. \textit{baseline}). 

We found that participants, including those in the baseline conditions, had some basic understanding of AI/ML concepts like false positives and false negatives.  According to Model 1 in Table \ref{tab:results}, participants in the baseline condition on average answered \textit{63.2\%} objective comprehension questions correctly. Using one of our interfaces increased participants' comprehension. Participants in the confusion matrix condition got \textit{70.9\%} questions correct; the difference between the confusion matrix and baseline conditions was not significant. Participants in the text view condition got \textit{74.7\%} questions correct; the difference between the text view and baseline conditions was significant (coef. = 0.115, \textit{p} < 0.05). A T-test showed no statistical difference between participants' performance in the two interface conditions. 

When it comes to algorithmic trade-offs, our interfaces significantly improved participants' comprehension. According to Model 2, participants in the baseline condition correctly answered \textit{36.7\%} of the objective comprehension questions on trade-offs.  Participants in the confusion matrix condition correctly answered \textit{55.8\%} of the questions; the difference between the confusion matrix and baseline conditions was significant (coef. = 0.191, \textit{p} < 0.01). Participants in the text view condition correctly answered \textit{65.6\%} of the questions; the difference between the text view and baseline conditions was significant (coef. = 0.289, \textit{p} < 0.01). A T-test between the two interface conditions did not show significant difference. 

We also examined the impact of using our interfaces on participants' subjective evaluation of their understanding of algorithmic trade-offs (Model 3). There was no significant difference among participants in the two conditions.

\begin{table}[b]
\small
\center
\begin{tabular}{|l|cc|cc|cc|}
\hline
\multicolumn{1}{|c|}{}          & \multicolumn{2}{c|}{\begin{tabular}[c]{@{}c@{}}Obj. Comp.\\ of Algorithmic \\ Concepts\end{tabular}} & \multicolumn{2}{c|}{\begin{tabular}[c]{@{}c@{}}Obj. Comp.\\ on Algorithmic \\ Trade-Offs\end{tabular}} & \multicolumn{2}{c|}{\begin{tabular}[c]{@{}c@{}}Sub. Evaluation\\ on Understanding \\ Trade-Offs\end{tabular}} \\ \hline
\multirow{2}{*}{}               & \multicolumn{2}{c|}{Model 1}                                                                                   & \multicolumn{2}{c|}{Model 2}                                                                                     & \multicolumn{2}{c|}{Model 3}                                                                                     \\ \cline{2-7} 
                                & Coef.                                                   & S.E.                                                 & Coef.                                                    & S.E.                                                  & Coef.                                                    & S.E.                                                  \\ \hline
Intercept                       & 0.632 **                                                & 0.031                                                & 0.367 **                                                 & 0.036                                                 & 5.264 **                                                 & 0.129                                                 \\ \hline
IsCMView                       & 0.077                                                      & 0.045                                                & 0.191 **                                                 & 0.052                                                 & 0.209                                                    & 0.189                                                 \\
IsTextView                      & 0.115 *                                                 & 0.046                                                & 0.289 **                                                 & 0.053                                                 & 0.236                                                    & 0.193                                                 \\ \hline
\multicolumn{1}{|c|}{R-Squared} & \multicolumn{2}{c|}{0.023}                                                                                     & \multicolumn{2}{c|}{0.099}                                                                                       & \multicolumn{2}{c|}{0.007}                                                                                       \\ \hline
\end{tabular}
\caption{Results of participants' objective comprehension and subjective evaluation of the algorithm.  *\textit{p} < 0.05, **\textit{p} < 0.01.}
\label{tab:results}
\end{table}

\subsubsection{Enabled Participants to Select the Models They Want}

Both interfaces (confusion matrix view and text view) enabled participants to view and select models ranging over a spectrum of trade-offs. We explicitly asked participants whether they think the interfaces helped them identify models that represented their values. 
Participants in both interface conditions reported high ratings. The average ratings for the confusion matrix condition and text view condition were 5.54 and 5.64 respectively on a 7-Likert scale (the difference was not statistically significant).

\begin{figure}[!ht]
\centering
\includegraphics[width=.8\columnwidth]{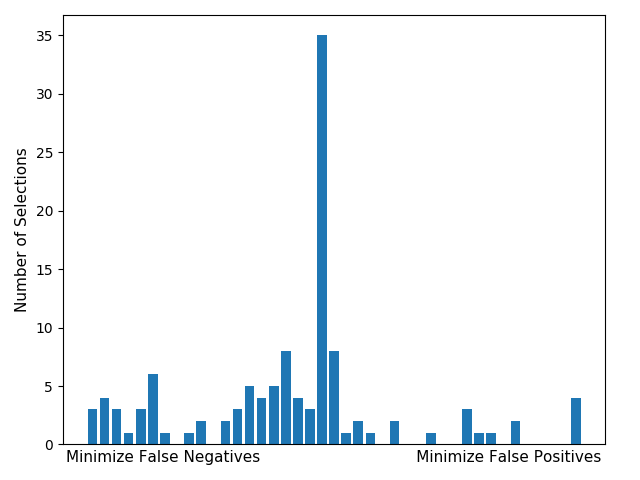}
  \caption{The distribution of participants' model selections with respect to trade-offs between false negatives and false positives. 
  }
  \label{fig:model_selection1}
\end{figure}

\begin{figure}[!ht]
\centering
\includegraphics[width=.8\columnwidth]{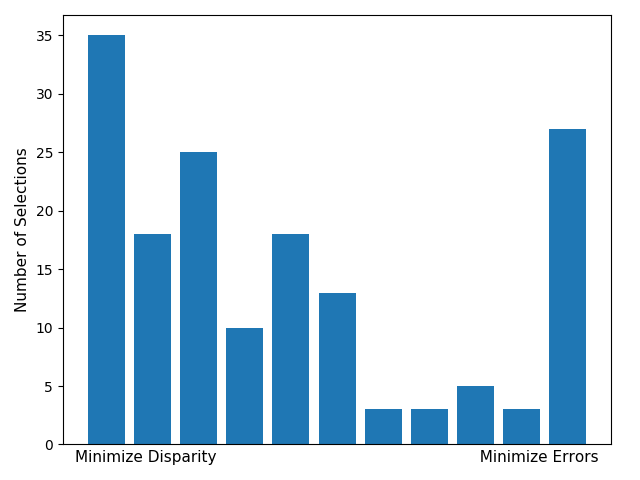}
  \caption{The distribution of participants' model selections with respect to trade-offs between disparity and prediction errors. 
  }
  \label{fig:model_selection2}
\end{figure}

Interestingly, we found great diversity in people's model selection: different people had different preferences for the type of outcomes and trade-offs they considered acceptable. Figure \ref{fig:model_selection1} and \ref{fig:model_selection2} show the distribution of participants' model selection. 

Figure \ref{fig:model_selection1} concerns the trade-off between false positives and false negatives. It shows that many participants tended to \textit{balance} the two types of errors in their model selections. 29.4\% of participants selected the model in the middle, which minimized the overall errors. Among those who did not select a ``balanced'' model, more participants selected models on the ``reducing false negatives'' side than models on the ``reducing false positives'' side. That is, they prioritized releasing defendants who might re-offend over retaining defendants who would not re-offend. 

Figure \ref{fig:model_selection2} concerns the  trade-off between overall errors and disparity. It shows a different pattern, a more bimodal distribution of preferences. 21.8\% of participants selected a model that minimized disparity, and 16.8\% selected a model that minimized overall errors . Among those who selected a model with some balance between the two goals, more participants preferred reducing disparity over reducing overall errors.

This suggests intriguing opportunities and future work on how to aggregate different individuals' opinions on ``what the best model is'', which we  discuss below.

\subsubsection{Making Trade-offs Transparent Swayed Participants' Trust}

We also measured the change in participants' perceived trust of algorithmic prediction before and after they explored the algorithm and trade-offs.  47.4\% participants in the two interface conditions changed their perceived trust, with a nearly even split between those who increased and those who decreased their trust. In the baseline condition, participant directly proceeded into questionnaires. Simply answering the questions about algorithmic trade-offs changed 30\% of participants' perspectives (the difference between the interface conditions and the baseline condition is significant, \textit{p} < 0.01). 

Participants explained their reasons for changing their perceived trust toward algorithmic prediction in an open-ended follow-up question. Analysis of their responses revealed some insights into their reasons.  

Many participants gained trust because our interfaces educated them about the algorithm itself and the inherent trade-offs in algorithms like this. One participant made this explicit:
    \textit{``Using the tool helped me understand the algorithm and the results of changing the aggressiveness and disparity parameters...''} 
Our interfaces also made it easier to see the prediction results. One participant said that it \textit{``makes it much easier to see how many false predictions the model can make''}.


On the other hand, the ability to tune the algorithm and see different prediction outcomes led some participants to doubt the algorithm's reliability and thus reduce their trust. As one participant said:  \textit{``After learning more about the algorithm, it seems that the parameters can be adjusted to create almost any type of results desired by researchers.''}.






Another interesting observation is that participants had very different expectations about algorithmic accuracy, and this affected their trust. For example, one participant increased trust saying that \textit{``an accuracy of around 70\% is fairly good''}. On the other hand, another participant commented that the algorithm is  \textit{``less trustful as this is a large error rate''}.

Overall, our first evaluation showed that our interfaces were effective at helping novice users comprehend and navigate trade-offs. In addition, our results suggested that people have heterogeneous preferences for fairness-accuracy trade-offs, and diverse perspectives on the trustworthiness of AI systems. This opens up new challenges and opportunities in building solutions with AI algorithms that take into account the diversity of human preferences.

To understand how our method might help expert users of a recidivism prediction tool (e.g., judges, lawyers, and policymakers), we recruited and conducted in-depth interviews with domain experts in this area.

\section{Evaluation 2: Expert Study}


\subsection{Recruitment, Procedure, and Analysis}

We recruited 6 experts who have extensive experiences in criminal justice system. They had backgrounds ranging from criminal law, justice, and public policies and statistics. Participation was voluntary and uncompensated. Participants were 2 females and 4 males based in the U.S. We began our recruitment at a conference on fairness in machine learning, and utilized a snowball sampling technique to identify more participants.  We conducted semi-structured interviews, and each lasted on average 30 minutes. Because our participants were remotely located, we used zoom.us and asked our participants to share screen. By doing so, we were able to observe how participants were interacting with our interface in real time and ask follow-up questions. 


We explained the context and goals of the study, asked for consent to record, and then gave participants time to explore the interface. After some exploration and clarification questions, participants were asked to think out loud their thought process and complete tasks such as describing trade-offs indicated by the interface, and identifying a model given a specific desired property. 
Participants were provided both interfaces (confusion matrix view and text view). In the interview, we asked our expert participants to envision how real users might use our interfaces in the real-world environment. 





All the participants quickly learned how to use interface, accurately identified the trade-offs, and answered the objective comprehension questions correctly without difficulty. 

To analyze the interview transcripts, we adopted Charmaz’ approach\cite{charmaz2014constructing} to grounded theory, so that prior ideas and theory could be considered during analysis. The specific coding process is as follows: members of the research team transcribed all 6 interviews from 3.5 hours of recorded audio, and open coded all transcripts. A total of 7 themes emerged from a series of immersive meetings where we discussed and clustered codes. We only reported selected themes that are relevant to the research questions of this paper.


\subsection{Results}

\subsubsection{Envision the Use in Practice}


Our expert participants believed that our interface is a great tool to reveal trade-offs, and encourage real users to think about trade-offs and consequences. 
    \textit{``I think you're onto something in that, it's useful to have.. They (different models) have these different trade-offs and consequences.' '' (P1)}


Furthermore, our interfaces made the ethical trade-off decisions transparent. 
    \textit{``The interface can make that trade-off clear, but it cannot help with the normative questions.  Making it clear can clarify people to think about it, where they want the line to be drawn at these normative questions.''}

Our methods also creates the opportunity for users to express their perception of different misclassfied cases, and even provide different weights to different misclassified cases. 
    \textit{``It's up to the stakeholders to decide.. they'll say to me false positive is three times worse than a false negative...''(P2)}

Some expert participants suggested that our interface can be useful at the final stage of model selection in the development of the algorithmic system:
    \textit{``...it would be like, people had already decided that one... the final thing of this would be to pick a threshold, or to access its predictive accuracy is suitable for deployment at all...'' (P3)}


In addition, some experts believed this is also a tool for educating the public. As P3 commented that 
    \textit{``I guess I could less imagine. like policy makers sitting in a room and using this, like more imagine the general public being interested in how this sort of thing gets done, like an instructive tool.'' (P3)}

\subsubsection{Interface Preference}
There was no general preference for one interface over the other. Some participants liked the text view more because they thought it was simpler and cleaner. As P3 said:
    \textit{``I like the text view, easy to observe info. confusion matrix draws too much attention ...
    I think it [text view] looks really clean and I think it's nice to see those numbers like, written out, I mean that's me though, like I like numbers, I like the `incorrect, correct' sort of labeled this way it's like obvious.''}

On the other hand, some participants liked the interactive visual feature of the confusion matrix view. As P5 said:
    \textit{``I think the first one [confusion matrix view] was a lot niftier (laughs.) ... I'm more intuitive. See the things changing visually... it's hard to see what how the trade-offs are happening as I move these without the little dots.''}

\section{Discussion}


In this research, we proposed a method that lets designers and users directly see the trade-offs in a range of AI prediction models. The "proof-of-concept" case study demonstrates that the method is promising to help both novice users and domain experts comprehend, interpret, navigate, and reflect on the algorithmic trade-offs. Specifically, the interfaces developed in the case study let people explore multiple points in the design space of recidivism prediction models and identify a model consistent with their values. 

Our findings also hold promise for designers who need to make judgments about these algorithms that can strongly impact user experiences. Our approach has the potential to allow for more fluid cross disciplinary development of algorithmic systems. However, we did not test our method and interfaces with designers and developers in a real design and development scenarios. In future work, we will organize workshops with designers and developers to understand how the tool can facilitate better communications in multi-disciplinary team.

We found that people have heterogeneous preferences about fairness-accuracy trade-offs in their model selections. 
This opens up a new challenge: how can we help users with different preferences negotiate and select a final model? One promising technical approach is to draw techniques from social choice theory and to develop mechanisms that elicit preferences from individual stakeholders and select models based on scoring rules (e.g. the Borda count~\cite{borda,Lee2018WeBuildAI}).  An alternative approach is to develop social mechanisms and user interfaces to facilitate discussions among groups of stakeholders, enabling them to reason about different fairness and accuracy measures, express priorities and acceptable trade-offs, and negotiate with each other and find appropriate models.

One surprising finding was that making trade-offs transparent changed people's trust of the algorithms in both positive and negative dimensions. This suggests that in future work we could actually seek to understand if a particular interface design increases or decreases trust in the technology. 
Another way to interpret our finding is that the knowledge that our tool offers helped people make informed decisions on whether they should trust the algorithms or not, which is critical for high-stake contexts. According to Simmel \cite{simmel1950sociology}, there is no need to trust if people have total knowledge of the other party (algorithms in our context). Trust is also not a rational choice if people do not have any knowledge of the other party. Trust exists when people have some knowledge about the other party \cite{simmel1950sociology}. We believe this opens up new opportunities to further understand how to design more trustworthy algorithmic systems.

We evaluated our method in the context of predicting recidivism because recidivism prediction tools are assisting humans to make consequential decisions and have received much attention from the public and research communities \cite{HowWeAnalyzedCOMPAS,MachineBias}. Prior work in fair machine learning has shown an inherent trade-off between fairness and accuracy for almost all prediction tasks. For example, Kearns et al have shown similar trade-offs (Pareto curves) between fairness measures and accuracy in many datasets, including the “lawschool” and “communities and crime” datasets \cite{empirical_gerry}. Thus, we believe our overall approach can be generalized to explain trade-offs between accuracy- and fairness-related measures in other contexts. However, different interfaces may be suitable for different contexts, as is the case with any interface design. 

Finally, one line of future work we are pursuing is to create an ``authoring tool'' to let designers (who may not have programming and algorithm development experience) create their own visualizations of trade-offs between different accuracy and fairness measures for the algorithmic systems they are designing.

\section{Conclusion}
In this study, we developed a method to explain trade-offs between design goals in the machine learning algorithm. 
We evaluated the method in the context of predicting criminal defendants' likelihood to re-offend through a large-scale online experiment and in-depth interviews with domain experts. Our results suggest our method is promising in helping designers and users comprehend, navigate, and manage trade-offs. 

\section{Acknowledgement}
This work was supported by the National Science Foundation (NSF) under Award No. IIS-2001851 and No. IIS-2000782, the NSF Program on Fairness in AI in collaboration with Amazon under Award No. IIS-1939606, and the JP Morgan Faculty Award.



\balance{}

\bibliographystyle{SIGCHI-Reference-Format}
\bibliography{proceedings.bbl}

\end{document}